\documentclass[aps,twocolumn,groupedaddress]{revtex4}

\usepackage{graphicx}%
\usepackage{dcolumn}
\usepackage{amsmath}
\usepackage{latexsym}

\newtheorem{theorem}{Theorem}

\begin{document}

\title{Local Distinguishability of Multipartite Orthogonal Quantum States}

\author{Jonathan Walgate}
\email[Correspondence to: ]{jon.walgate@qubit.org}
\author{Anthony J. Short}
\author{Lucien Hardy}
\author{Vlatko Vedral}
\affiliation{Centre for Quantum Computation, Clarendon Laboratory, Parks Road, Oxford OX1 3PU, United Kingdom}

\date{\today}

\begin{abstract}
We consider one copy of a quantum system prepared in one of two orthogonal pure states,
entangled
or otherwise, and distributed between any number of parties.  We demonstrate that it is
possible to identify which of these two states the system is in by
means of local operations and
classical communication alone. The protocol we outline is both completely
reliable and completely general - it will correctly distinguish
any two orthogonal states 100\% of the time.
\end{abstract}

\maketitle

\section{Introduction}
\label{sec:Intro}
Pure quantum states may only be perfectly distinguished from one
another when they are orthogonal. That is, a
state $\left| \psi \right\rangle $ may be reliably distinguished
from another, $\left| \phi \right\rangle $, only if $\left\langle
\psi |\phi \right\rangle =0$. We will show that if $\left\langle
\psi |\phi \right\rangle =0$ for given $\left| \psi\right\rangle$
and $\left| \phi\right\rangle$, then $\left| \psi \right\rangle $
may always be distinguished from $\left| \phi \right\rangle $ by
means of local operations and classical communication (LOCC).
This may be surprising, since quantum systems can
encode information that may only be extracted by analyzing the
system \emph{as a whole}. This
well-known phenomenon - entanglement - forms the basis of many
recently proposed quantum schemes,
such as cryptography\cite{crypt1, crypt2, crypt3} computation\cite{comp} and enhanced
communication\cite{comm}.
A tempting interpretation is that ``entangled
information'' can only be uncovered using global measurements upon
the system as a whole. But this is not the case - in our very
general situation local measurements, sequentially dependent upon
classically communicated prior measurement results, suffice to
identify orthogonal entangled quantum states.

Schemes for distinguishing
between a set of quantum states, both pure and mixed have been considered by
various authors \cite{helstrom, holevo, fuchs, sausage, gottesman, koashi}. Closely related to the
present paper is the work of Bennett et al \cite{sausage} who showed that there exist sets of
orthogonal product states  that cannot be distinguished by LOCC.

Alice and Bob each hold part of a quantum system, which occupies one of
two possible orthogonal quantum states $\left|
\psi \right\rangle $ and $\left|
\phi \right\rangle $.
Alice and Bob know the precise form of $\left| \psi \right\rangle $ and $\left| \phi
\right\rangle $, but have no idea which of these possible
states they actually possess: they will have to
perform some measurements to find out. A global measurement would suffice,
but alas Alice and Bob cannot afford to meet up. Fortunately for them,
they are on speaking terms, as one phone call is all they require. This
situation, LOCC, is of primary relevance to most
applications of entanglement.

The strategy Alice and Bob adopt
is simple. They can always find a basis in which the two orthogonal
states can be represented
\begin{eqnarray}
\left| \psi \right\rangle &=&\left| 1\right\rangle _{A'}\left| \eta
_{1}\right\rangle _{B}+\cdots +\left| l\right\rangle _{A'}\left|
\eta _{l}\right\rangle _{B}  \label{final} \\ \left| \phi
\right\rangle &=&\left| 1\right\rangle _{A'}\left| \eta
_{1}^{\perp }\right\rangle _{B}+\cdots +\left| l\right\rangle
_{A'}\left| \eta _{l}^{\perp }\right\rangle _{B}  \nonumber
\end{eqnarray}
where \{ $\left| i\right\rangle_{A'}$ for $i=1$ to $l$\} form some orthogonal
basis set for Alice, $\{ \left| \eta _{1}\right\rangle_{B} ,\cdots ,\left| \eta _{l}\right\rangle_{B}\}$
are not normalized, and $\left| \eta _{i}^{\perp
}\right\rangle _{B}$ is orthogonal to $\left| \eta _{i}\right\rangle_{B}$. Alice simply measures her part of the system in such a basis, and
communicates the result, $i$, to Bob. Bob
then has an easy task - he may distinguish locally between $\left| \eta
_{i}\right\rangle _{B}$ and $\left| \eta _{i}^{\perp
}\right\rangle _{B}$ and thereby know which state he and Alice
shared to begin with.
\section{Matrix Representation of Possible States}
\label{sec:2}
Alice and Bob start out knowing the precise form of two states
that might correspond to their shared quantum system. These two possible
states, $\left| \psi \right\rangle $ and $\left| \phi
\right\rangle $, are orthogonal, so that $\left\langle \psi |\phi
\right\rangle =0$. We can represent them in the following,
entirely general way:
\begin{eqnarray}
\left| \psi \right\rangle &=&\left| 1\right\rangle _{A}\left| \eta
_{1}\right\rangle _{B}+\cdots +\left| n\right\rangle _{A}\left|
\eta _{n}\right\rangle _{B}  \label{initial} \\ \left| \phi
\right\rangle &=&\left| 1\right\rangle _{A}\left| \nu
_{1}\right\rangle _{B}+\cdots +\left| n\right\rangle _{A}\left|
\nu _{n}\right\rangle _{B}  \nonumber
\end{eqnarray}
where $\{ \left| 1\right\rangle _{A},\cdots ,\left| n\right\rangle _{A}\}$ form
an orthonormal basis set for Alice, and the vectors $\{\left| \eta
_{1}\right\rangle _{B},\cdots ,\left| \eta
_{n}\right\rangle _{B}\}$ and $\{\left| \nu
_{1}\right\rangle _{B},\cdots ,\left| \nu
_{n}\right\rangle _{B}\}$ are not
normalized and also not necessarily orthogonal. Alice and Bob can express the vectors
$\{\left| \eta
_{1}\right\rangle _{B},\cdots ,\left| \eta
_{n}\right\rangle _{B}\}$ and $\{\left| \nu
_{1}\right\rangle _{B},\cdots ,\left| \nu
_{n}\right\rangle _{B}\}$ as a
superposition of a set of arbitrary basis vectors \mbox{$\{ \left| 1\right\rangle _{B},\cdots ,\left| m\right\rangle _{B}\}$}
in Bob's space
\begin{equation}
\left| \eta_{i}\right\rangle _{B}=\sum_{j} F_{ij} \left| j\right\rangle _{B}
\, , \;\;
\left| \nu_{i}\right\rangle _{B}=\sum_{j} G_{ij} \left| j\right\rangle _{B}
\label{A,B matrices}
\end{equation}
where the elements $F_{ij}$ an $G_{ij}$ form two $n\times m$ matrices $F$
and $G$.
These matrices preserve all the information Alice and Bob hold
about states $\left| \psi\right\rangle$ and $\left| \phi\right\rangle$.
Because of the way they are constructed, the
matrix $FG^{\dagger}$ takes the following form:
\begin{equation}
FG^{\dagger}=\left(
\begin{array}{ccc}
\langle \nu _{1}|\eta _{1}\rangle & \cdots & \langle \nu _{1}|\eta
_{n}\rangle \\ \vdots & \ddots & \vdots \\ \langle \nu _{n}|\eta
_{1}\rangle & \cdots & \langle \nu _{n}|\eta _{n}\rangle
\end{array}
\right)  \label{AB}
\end{equation}
We can see this is the case by inspection, because \mbox{$\left\langle \nu_{i} |\eta_{j} \right\rangle = \sum_{k=1}^{n}
F_{jk}G_{ik}^*$}.
The matrix $FG^{\dagger}$ encapsulates a great deal of
significant information for Alice and Bob about the relationship
between the states $\left| \psi\right\rangle$ and $\left| \phi\right\rangle$. Since we know by the
conditions of the problem that $\langle \phi |\psi \rangle =0$,
we know that
\begin{equation}
\langle \phi |\psi \rangle =\sum_{i=1}^{n} \langle \nu _{i}|\eta _{i}\rangle = {\rm Trace}(FG^{\dagger})=0 \label{orthogonality}
\end{equation}
But the $FG^{\dagger}$ matrix holds more information than the
simple fact of the states' orthogonality. It also encodes the key
to distinguishing between these two possible states. Alice plans
to distinguish $\left| \psi\right\rangle $ and $\left| \phi\right\rangle$ by
finding some basis - any basis - in which she can describe her part such that the states $%
\left| \psi\right\rangle $ and $\left| \phi\right\rangle $ take the more restricted form of (\ref{final}).
Alice must choose her $\{ \left| 1\right\rangle _{A},\cdots ,\left| n\right\rangle _{A}\}$ basis carefully such that no matter what
result $\left| i\right\rangle _{A}$ she obtains, Bob can surely
distinguish between his possible states. This means that for all $i$ , $%
\left| \nu _{i}\right\rangle $ must be orthogonal
to $\left| \eta _{i}\right\rangle $. Thus we can write down our distinguishability
criterion:
\begin{equation}
\forall i \qquad \langle \nu _{i}|\eta _{i}\rangle =0
\label{distinguishability}
\end{equation}
In other words, in our matrix representation, we require the
diagonal elements of $FG^{\dagger}$ to be
zero. Alice can alter the form of $FG^{\dagger}$ by changing the basis in
which she describes and measures her system. She has a
great deal of choice in this regard: any
orthogonal basis set spanning her space will provide a
description of form (\ref{initial}), and thus some matrix
$FG^{\dagger}$ of form (\ref{AB}). When she changes her orthonormal
basis set, this
changes the form of the matrices $F$ and $G$, and thus changes the form of
$FG^{\dagger}$. In fact, unitary transformations of Alice's measurement
basis map to the conjugate unitary transformations upon
$FG^{\dagger}$.
\begin{theorem}
A unitary transformation $U^{A}$ upon Alice's measurement
basis will transform the matrix $FG^{\dagger}$ to $U^{A\ast }(FG^{\dagger})U^{A\ast \dagger }$.
\end{theorem}

Proof: From (\ref{initial}),
$\left| \psi\right\rangle =\sum_{i}\left| i\right\rangle _{A}\left| \eta
_{i}\right\rangle _{B}$. Alice's unitary transformation acts
thus: $\left| i\right\rangle _{A}=\sum_{j}U_{ij}^{A\dagger }\left|
j^{\prime }\right\rangle _{A}$. From (\ref{A,B matrices}) it follows that, in Alice's
new basis $\{\left| 0^{\prime }\right\rangle _{A},\cdots ,\left| n^{\prime }\right\rangle
_{A}\}$:
\begin{equation}
\left| \psi\right\rangle =\sum_{ijk}U_{ij}^{A\dagger }\left| j^{\prime }\right\rangle
_{A}F _{ik}\left| k\right\rangle _{B}
\end{equation}
For true generality, we consider Bob might assist
Alice by unitarily rotating his basis by $U^{B}$. We therefore
write \mbox{$\left| k\right\rangle _{B}=\sum_{l}U_{kl}^{B\dagger
}\left| l^{\prime }\right\rangle _{B}$}, giving
\mbox{$\left| \psi\right\rangle =\sum_{ijkl}\left| j^{\prime }\right\rangle _{A}\left|
l^{\prime }\right\rangle _{B}U_{ij}^{A\dagger }F
_{ik}U_{kl}^{B\dagger }$}.
Since $U_{ij}^{A\dagger }=U_{ji}^{A\ast } $, we can rewrite this
as
\begin{equation}
\psi =\sum_{ijkl}\left| j^{\prime }\right\rangle _{A}\left|
l^{\prime }\right\rangle _{B}U_{ji}^{A\ast }F
_{ik}U_{kl}^{B\dagger }.
\end{equation}
By analogy with (\ref{initial}) and (\ref{A,B matrices}), this
means that in the new basis of description, we have a new matrix
$F^{\prime }$ where
$F _{ik}^{\prime }=\sum_{jl}U_{ji}^{A\ast }F
_{ik}U_{kl}^{B\dagger }$. Under unitary basis rotations by Alice
and Bob, our matrices $A$ and $B$ undergo the curious
transformations
\begin{equation}
F^{\prime }=U^{A\ast }FU^{B\dagger }\mbox{}\mbox{}\mbox{},\mbox{}\mbox{}\mbox{} G^{\prime
}=U^{A\ast }GU^{B\dagger } \label{A,B transformations}
\end{equation}
This means that the object of our interest, the $FG^{\dagger}$
matrix encoding information about the relationship
\emph{between} the states, will transform as...
\begin{eqnarray}
F^{\prime }G^{\prime \dagger }=(U^{A\ast }FU^{B\dagger })\left(
U^{A\ast }GU^{B\dagger }\right) ^{\dagger } \nonumber \\ =U^{A\ast }FU^{B\dagger
}U^{B}G^{\dagger }U^{A\ast \dagger } \nonumber \\ =U^{A\ast }(FG^{\dagger
})U^{A\ast \dagger } \qquad \Box \label{AB transformation}
\end{eqnarray}
Bob's unitary rotation $U^{B}$ drops out, as
rotations in his basis will not affect the overlaps
$\langle \nu _{i}|\eta _{j}\rangle$ that make up $FG^{\dagger}$.

If $U^{A}$ is unitary, then so is $U^{A\ast }$. Alice
can find a basis of form (\ref{final}), and thereby satisfy our
distinguishability criterion (\ref{distinguishability})
, \emph{if and only if} there exists a unitary matrix $%
U=U^{A\ast }$ such that $U(FG^{\dagger})U^{\dagger }$ is a
``zerodiagonal'' matrix. (A matrix whose diagonal elements are all zero.)
A proof that such a unitary matrix always
exists constitutes a proof that two orthogonal quantum states can
always be distinguished.
\section{Matrix proof of $\left| \psi\right\rangle ,\left| \phi\right\rangle $ distinguishability}
\label{sec:3}
Unitary transformations upon Alice's measurement basis
translate into (conjugated) unitary transformations upon her specific $%
FG^{\dagger}$ matrix. If she can find a unitary rotation that
converts this matrix into zerodiagonal form, she can ensure Bob
will be able to distinguish between states $\left| \psi\right\rangle $ and $\left| \phi\right\rangle $.

We first prove such a rotation always exists in the two-dimensional case, and then show how
Alice may use a finite sequence of such $2\times2$
transformations to zerodiagonalize any traceless $n\times n$
matrix.
\subsection{Two-dimensional case}
\label{sec:3.1}
\begin{theorem}
Let $M$ be the wholly general $2\times 2$ matrix $\left(
\begin{array}{cc}
x & y \\ z & t
\end{array}
\right) $. There exists a $2\times 2$
unitary matrix $U$ such that the diagonal elements of \mbox{$UMU^{\dagger
}$} are equal.
\end{theorem}
Proof: Let $U=\left(
\begin{array}{cc}
\cos \theta  & \sin \theta e^{i\omega
}  \\ \sin \theta e^{-i\omega
} & -\cos \theta
\end{array} \right)$.

We need the diagonal elements of $UMU^{\dagger }$ to be equal.
This gives us the condition:
\begin{equation}
(x-t)\cos 2\theta +\sin 2\theta (ye^{-i\omega }+ze^{i\omega })=0
\end{equation}
The real and imaginary parts of this equation, can be solved for the angles $\omega$ and $\theta$:
\begin{equation}
\tan \omega =\frac{{\rm Im}(x-t){\rm Re}(z+y)-{\rm Re}(x-t){\rm Im}(z+y)}{%
{\rm Re}(x-t){\rm Re}(z-y)+{\rm Im}(x-t){\rm Im}(z-y)} \label{omega}
\end{equation}
\begin{equation}
\tan 2\theta =\frac{{\rm Re}(x-t)}{{\rm Re}(z+y)\cos \omega -{\rm Im}(z-y)\sin \omega}
\label{theta}
\end{equation}
The RHS of (\ref{omega}) is always real, and thus there will
always be an angle $\omega $ that satisfies the equation. Given a
definite $\omega $, we can always solve (\ref{theta}) for a definite
$\theta $ for the same reason. Thus for any $2\times 2$ matrix
$M$, there exists a $2\times 2$ unitary matrix that ``equidiagonalizes''
it. (Equalizes all its diagonal elements.) This completes the
proof $\Box$.

This mathematical result can be applied to the $2 \times 2$ dimensional case.
Since the $\left| \psi\right\rangle$ and $\left| \phi\right\rangle$ states are orthogonal, the corresponding
$FG^{\dagger}$ matrix
is traceless, in which case equidiagonalization constitutes
zerodiagonalization. Equations (\ref{omega}) and (\ref{theta}) therefore always pick out
a specific unitary transformation that will zerodiagonalize $FG^{\dagger}$. By measuring
in that basis, Alice and Bob can always distinguish between the two possible
orthogonal states of their system.
\subsection{ \; $2^k$ dimensional case}
\label{sec:3.2}
We want to consider all situations of greater dimensionality than
2, but we first concentrate on situations where Alice's
Hilbert space has $2^k$ dimensions (where $k$ is some positive integer). The $AB^\dagger$ matrix has
the same dimensionality, and will have $2^k \times 2^k$ elements.
Note that while this particular class of $FG^{\dagger}$ matrices
- those of dimension $2^{k}$ - may seem
limited, it includes all quantum states comprising sets of qubits.
In such cases, Alice can adopt a simple strategy to
equidiagonalize this potentially huge matrix in a relatively
small number of steps. We know from theorem 2 above that Alice
may unitarily rotate any two diagonal elements in her
$FG^{\dagger} $ matrix so that they become equal. By grouping the
diagonal elements into $2^{k-1} $ pairs, and equidiagonalizing each pair,
she can create $2^{k-1} $ equal pairs.

Both elements of an equal pair can then be individually made equal
to the elements of another equal pair, using only two $2
\times 2$ unitary transformations. Thereby, Alice can create
$2^{k-2}$ ``quartets'' of equal diagonal elements with just $2^{k-1} $ further $2\times 2$ unitary transformations.
By repeating this process $k$ times, Alice will set all the diagonal elements exactly
equal. If her $FG^{\dagger}$ matrix has $2^{k}$ diagonal elements, then $k\cdot
2^{k-1}$ elementary operations will serve to equidiagonalize it.
This satisfies Alice's requirements: since she knows that her
physical $FG^{\dagger}$ matrix is traceless, she knows that all
the diagonal elements $\langle \nu _{i}|\eta _{i}\rangle$ will be thereby set to
zero. Therefore Alice and Bob can distinguish the two orthogonal
states.
Of course, Alice need not physically enact each and every
separate $2\times 2$ unitary transformation.
A single $2^k \times 2^k$ unitary transformation will represent the product of all these
rotations, and finding this one transformation that equidiagonalizes
$FG^{\dagger}$ in one shot is a perfectly tractable problem for Alice to
solve.
\subsection{General case}
\label{sec:3.3}
The matrix $FG^{\dagger}$ will not, in general, be of size
$2^{k}\times 2^{k}$. Alice may nevertheless devise an approach that
is guaranteed to yield state equations of form (\ref{final}).
She needs to be inventive. Her
favored tactic so far - a sequence of pair-wise equalizations -
will converge upon the desired unitary matrix only in the infinite
limit. She can find a more elegant method, however.
The $2^{k}$ dimensional case is unproblematic, so if Alice can \emph{enlarge}
$FG^{\dagger}$ such that it achieves a dimensionality of a power of
two, she can solve her problem.

Such an enlargement represents an expansion of Alice's quantum system
into a Hilbert space of greater dimension. She must perform a SWAP operation to
transfer the state of her original quantum system $\mathcal{H}_{n}^{A}$
described by (\ref{initial}) to an $n$-dimensional subspace of a larger space, $\mathcal{H}_{l}^{A'}$,
where $l\geq n$ and $l=2^{k}$ for some integer $k$:
\begin{eqnarray}
\left| i\right\rangle_{A} \left|j \right\rangle_{A'} &\Longrightarrow& \left|j \right\rangle_{A} \left|i
\right\rangle_{A'} \;\;\mbox{when}\;\; i,j=1 \;\mbox{to}\; n \\
\setlength{\mathindent}{100pt}
\left| i\right\rangle_{A} \left|j \right\rangle_{A'} &\Longrightarrow& \left|i \right\rangle_{A} \left|j
\right\rangle_{A'} \;\;\mbox{otherwise} \nonumber
\end{eqnarray}
Since the size of $FG^{\dagger}$ is simply equal to the number of orthonormal
vectors in Alice's measurement basis, this operation expands it to size $l \times
l$. In her new basis, $\{ \left| 1\right\rangle _{A'},\cdots ,\left| l\right\rangle _{A'}\} _{A}$, Alice describes
the two possible states (\ref{initial}) thus:
\begin{eqnarray}
\left| \psi \right\rangle &=&\left| 1\right\rangle _{A'}\left| \eta
_{1}'\right\rangle _{B}+\cdots +\left| l\right\rangle _{A'}\left|
\eta _{l}'\right\rangle _{B} \\ \left| \phi
\right\rangle &=&\left| 1\right\rangle _{A'}\left| \nu
_{1}'\right\rangle _{B}+\cdots +\left| l\right\rangle _{A'}\left|
\nu _{l}'\right\rangle _{B}  \nonumber
\end{eqnarray}
Here, $\left| \eta_{i}'\right\rangle _{B}$ and $\left| \nu
_{i}'\right\rangle _{B}$ are new unnormalized vectors, but remain
describable in Bob's original basis $\{ \left| 1\right\rangle _{B},\cdots ,\left| m\right\rangle _{B}\}$.
Now her system has a convenient number of dimensions, Alice
proceeds as in Sec.~\ref{sec:3.2}. She will obtain and perform a measurement
guaranteeing Bob possesses one of two orthogonal states.

SWAP operations like these are physically unproblematic, and do not in any way derogate the entangled information
Alice shares with Bob. One physical realization of this
procedure requires just one ancilliary qubit. Alice introduces
this
qubit ``$Z$'', known to be in state $\left| 0 \right\rangle _{Z}
$ to her system, giving her state equations of form:
\begin{eqnarray}
\left| \psi \right\rangle = \left| 10\right\rangle _{AZ}\left| \eta _{1}\right\rangle _{B}+\cdots
+\left| n0\right\rangle _{AZ}\left| \eta
_{n}\right\rangle _{B}\\+\left| 11\right\rangle _{AZ}\left| \eta _{n+1}\right\rangle _{B}+\cdots
+\left| n1\right\rangle _{AZ}\left| \eta
_{2n}\right\rangle _{B} \nonumber
\end{eqnarray}
Since qubit $Z$ is in state $\left| 0\right\rangle
_{Z}$, we know all the unnormalized vectors $\left| \eta
_{n+i}\right\rangle _{B}$ have zero amplitude. This gives rise to the rather lop-sided
$FG^{\dagger}$ matrix, wherein $\{FG^{\dagger}\}_{ij}=0$ everywhere
that either $i>n$ or $j>n$. With this $FG^{\dagger} $ matrix, Alice's problems are over. Between
the numbers $n$ and $2n$ there lies a power of two. Thus
there is a sub-matrix of $FG^{\dagger} $ that includes all
$n$ non-zero terms, and just enough zero-valued terms to round
things out to the most convenient dimensionality. Alice can find
unitary manipulations on this sub-matrix that transform it, (and
thereby simultaneously transform the $FG^{\dagger} $ matrix as a
whole) into zero-diagonal form. She simply follows the procedure
outlined in Sec.~\ref{sec:3.2}, obtaining a finite sequence of
unitary transformations that, taken together, represent a single
rotation of her measurement basis.

This unlikely procedure is surprisingly efficient for
distinguishing $\left| \psi\right\rangle$ and $\left| \phi\right\rangle$. No matter what the
dimensionality of the problem, there is a solution after a finite
number of steps: a number of steps equal to $\frac{1}{2}\, l \log_{2}l$,
where $l$ is the expanded dimensionality. Through the use of this
SWAP operation
Alice can always accomplish perfect distinguishability with minimal effort.
\section{Further Generalizations}
\subsection{Multipartite states}
\label{sec:4.1}
We have considered only the bipartite case thusfar, but the
strategy used by Alice and Bob can also be deployed by any number
of people. States of tripartite form, for instance:
\begin{eqnarray}
\left| \psi \right\rangle &=&\left| \alpha _{0}\right\rangle
_{A}\left| \beta_{0}\right\rangle _{B}\left| \gamma
_{0}\right\rangle _{C}+\cdots +\left| \alpha _{n}\right\rangle _{A}\left|
\beta_{n}\right\rangle _{B}\left| \gamma _{n}\right\rangle _{C}
\\
\left| \phi \right\rangle &=&\left| \alpha _{0}^{\prime
}\right\rangle _{A}\left| \beta _{0}^{\prime }\right\rangle
_{B}\left| \gamma _{0}^{\prime }\right\rangle _{C}+\cdots +\left| \alpha _{n}^{\prime }\right\rangle _{A}\left|
\beta _{n}^{\prime }\right\rangle _{B}\left| \gamma _{n}^{\prime
}\right\rangle _{C}  \nonumber
\end{eqnarray}
can, when Alice swaps into a larger Hilbert space,
easily be represented thus:
\begin{eqnarray}
\left| \psi \right\rangle &=&\left| 0\right\rangle _{A'}\left|
\Gamma _{0}\right\rangle _{BC}+\cdots +\left|
l\right\rangle _{A'}\left| \Gamma _{l}\right\rangle _{BC} \\
\left| \phi \right\rangle &=&\left| 0\right\rangle _{A'}\left|
\Gamma _{0}^{\perp }\right\rangle _{BC}+\cdots
+\left| l\right\rangle _{A'}\left| \Gamma _{l}^{\perp
}\right\rangle _{BC} \nonumber
\end{eqnarray}
Alice simply behaves as before, and leaves Bob and Claire to
distinguish between the resulting bipartite orthogonal states. The problem
collapses to its original formulation, which we have already
solved. If $n$ people share the quantum system, performing a
series of $n-2$ such measurements will cascade their problem down to the bipartite case.
We can conclude that two orthogonal states of
any quantum system, shared in any proportion between any number
of separated parties can be perfectly distinguished.
\subsection{Multiple possible states}
\label{sec:4.2}
Our procedure distinguishes perfectly between two orthogonal
states, $\left| \psi \right\rangle $ and $\left| \phi
\right\rangle $. What if Alice and Bob must distinguish between
more than two orthogonal states? In general, this will not be
possible so long as Alice and Bob share only one copy of their
state. Whichever bases they perform sequential measurements in,
their binary outcomes may not perfectly distinguish between more
than two possibilities.

It is natural to quantify Alice and Bob's situation by asking
\emph{how many} copies of their state they require to perfectly
distinguish between it and the other possibilities. A detailed
analysis of this problem is beyond the scope of this paper.
Nevertheless, our basic procedure places an upper bound on the
number of copies required. $n$ possible orthogonal
states can be distinguished perfectly with $n-1$ copies.

Let us denote the possible states $\left| \psi _{i}\right\rangle
$. Alice
and Bob simply act on their first copy as if they were distinguishing $
\left| \psi _{0}\right\rangle $ and $\left| \psi_{1}\right\rangle
$. If the state they share happens to be either $
\left| \psi _{0}\right\rangle $ or $\left| \psi _{1}\right\rangle
$, then their measurement result will be a definite verdict in
favour of one or the other possibility. If they share instead some other $\left| \psi
_{i}\right\rangle$, since $\left\langle
\psi_{i} |\psi_{j} \right\rangle =\delta_{ij}$, Alice and Bob's measurement will
randomly decide upon $ \left| \psi
_{0}\right\rangle $ some of the time, and will seem to measure
$\left| \psi _{1}\right\rangle$ otherwise. A positive measurement
for $ \left| \psi _{0}\right\rangle $ is no guarantee of Alice and
Bob sharing that state, for all the other states (barring $\left|
\psi _{1}\right\rangle$) sometimes produce that result. What a
verdict for  $ \left| \psi _{0}\right\rangle $ does
show is that Alice and Bob definitely do not share $\left| \psi
_{1}\right\rangle$, which they would have detected with certainty.

Proceeding in this way, Alice and Bob can always use a single copy
of their state to exclude one possibility. After $n-1$ such
operations, they can have excluded $n-1$ states, and can thus
distinguish between $n$ possibilities. This represents an upper bound upon the number
of copies required for state distinction. Note that there are certainly sets
of orthogonal states that can be distinguished using less than
$n-1$ copies. An example are the four Bell-states, where
only two copies will suffice.
\section{Conclusion}
\label{sec:5}
We have proved that any two orthogonal quantum states shared
between any number of parties may be
perfectly distinguished by local operations and classical
communication. Since orthogonal states are the only perfectly
distinguishable states, this means that all pairs of
distinguishable states are distinguishable with LOCC - global
measurements are never required. Whether
non-orthogonal states may also be optimally
distinguished in this way remains an open question.

We would like to acknowledge the support of Hewlett-Pacard.
JW and AJS thank the UK \mbox{EPSRC} (Engineering
and Physical Sciences Research Council), and LH thanks the Royal
Society for funding this research.

\end{document}